\documentclass{osa-article}

\RequirePackage[utf8]{inputenc}
\RequirePackage[english]{babel}

\journal{josab}

\articletype{Research Article}

\begin{document}

\title{Tutorial on the analytical calculation of optical forces on spherical particles in optical tweezers}

\author{Antonio Alvaro Ranha Neves,\authormark{1,*} and Carlos Lenz Cesar\authormark{2}}

\address{\authormark{1}Centro de Ciências Naturais e Humanas, Universidade Federal do ABC (UFABC), Santo André - São Paulo, 09.210-170, Brazil\\
\authormark{2}Universidade Federal do Ceara, Departamento de Fisica, Fortaleza, CE, Brazil}

\email{\authormark{*}antonio.neves@ufabc.edu.br} 



\begin{abstract}
Arthur Ashkin was awarded the 2018 Nobel prize in physics for the invention of optical tweezers. Since the first publication in 1986 Optical Tweezers have been used as a tool to measure forces and rheological properties of microscopic systems. For the calibration of these measurements, the knowledge of the forces is fundamental. However, it is still common to deduce the optical forces from assumptions based on the particle size with respect to the trapping laser wavelength. This show the necessity to develop a complete and accurate electromagnetic model that does not depend on early approximations of the force model. Furthermore, the model we developed has several advantages, such as morphology--dependent resonances, size dependence for large spheres, and multipole effects from smaller particles, just to name a few. In this tutorial, we review and discuss the physical modeling of optical forces in optical tweezers, which are the resultant forces exerted by a trapping beam on a sphere of any size and composition.
\end{abstract}


\section{Optical trapping}
Optical trapping is a modern tool that employs a structured laser beam to manipulate and investigate cellular-scale physio-chemical phenomena. A few of the advantages of the trapping technique are that it exerts no physical contact and exhibits a very low stiffness, which makes it an ideal tool for soft-matter research. Moreover, the trapping beam can be split or multiplexed into several traps. During the last couple of decades, optical trapping has been greatly perfected owing to instrumental development and calibration techniques. Among the different optical trapping techniques, the single beam gradient optical trap is the prevailing one. This trap is also referred to as optical tweezers, a term coined by the technique´s pioneer Arthur Ashkin \cite{Ashkin1970,Ashkin1986}, and acknowledged with a shared Noble prize in physics in 2018. Ashkin's book \cite{Ashkin2006}, a collection of early articles with commentaries, illustrates the early years of optical tweezers research, along with some of the related modeling briefly mentioned in this tutorial.

The physical principle behind optical trapping lies in, the momentum of the electromagnetic field of the trapping beam, as stated by Maxwell. When this beam is incident upon an object, it is scattered elastically upon their encounter, resulting in a net change to the rate of linear momentum. Because of the conservation of linear momentum, a force is exerted on the object by the light beam. To obtain an intuitive idea of the force that can be exerted by light, the light's momenta is related to de Broglie relation from modern physics, which indicates a minuscule momentum per photon, resulting in a minuscule force, comparable to the weight of cellular-sized objects. For example, a 5 $\mu$m water droplet weight just as much as the force imparted on a mirror by a 1mW laser pointer.

To achieve a significant change in the linear momentum, a highly focused beam is required. In standard optical tweezers, this trapping beam is a single highly focused Gaussian profile laser beam. The trapped object is generally a dielectric microsphere, for which forces can be calibrated and employed as a force transducer to determine optical forces in other non-spherical samples within the same medium. 

\subsection{Optical force models}

Early theoretical modeling for optical forces conducted as early as 1909 by Debye, showing that the force on a sphere can be described in terms of extinction and scattering cross sections \cite{Debye1909,Kerker1969}. Debye considered a linearly polarized plane wave incident on a homogeneous spherical particle.

To theoretically determine these optical forces, initial assumptions are set based on the microsphere size with respect to the beam wavelength. For a small object that is much smaller than the beam wavelength, the object behaves like an electric dipole, in agreement with Rayleigh scattering. Hence, a simple description is established for optical forces acting on an electromagnetic dipole, and this is known as the Rayleigh regime \cite{Harada1996}. However, for objects greater than the wavelength, the trapping beam is treated as rays. The optical force is thus described in terms of geometrical optics, considering multiple reflections of the rays in the trapped microsphere to determine the net change in momentum \cite{Ashkin1992,Gussgard1992}. However, in the case of most experiments, the trapped object diameter is within the order of the laser wavelength, and the above mentioned approximations fail. A complete classical electromagnetic (EM) wave theory is needed to model optical forces in this intermediary regime, known as the Mie regime, although the Mie theory is applicable to spheres of any size. This analytical EM approach has been avoided mainly owing to its troublesome calculations; however, after some algebraic derivations, simple expressions for accurate optical forces, valid for any object size, can be obtained.

The complete EM approach allows to determine the morphology--dependent resonance, or whispering gallery modes, which have been observed in optical traps and are absent in the approximate size regime descriptions \cite{Ashkin1977,Fontes2005,Neves2006OE}. The early studies that considered the complete electromagnetic approach are those of Mie \cite{Mie1908} and Lorenz \cite{Lorenz1890}. Historical reviews of the initial developments and contributions are provided by Logan \cite{Logan1965} and Kerker \cite{Kerker1969}. The Lorenz-Mie theory is an analytical method designed for plane-wave scattering by a spherical particle, which is not applicable to optical tweezers that require highly focused beams. Therefore, a generalization for the scattering of generic beams by Mie sized particles led to the development of the generalized Lorenz-Mie theory (GLMT)\cite{Gouesbet1982,Gouesbet1988,Gouesbet2017}. In the GLMT framework, the incident and scattered beams are expressed in terms of vector spherical wavefunctions, whose expansion coefficients describe the applied beam configuration.

Once the vector scattered fields are known, we can, therefore, proceed with the Maxwell stress tensor (MST) approach to our problem and determine the net force or torque the beam exerts on the microsphere. This approach was first performed by Kim and Lee \cite{Kim1983}, where they presented a vector theory for optical forces acting on a sphere by a Gaussian beam described from a complex-source-point method.  The treatment relied on expanding the fields in vector spherical harmonics, much like GLMT. Subsequently and in a similar manner, Marston and Crichton \cite{Marston1984}, presented an \textit{ab initio} calculation of the radiation torque on an isotropic sphere illuminated by a circularly-polarized plane electromagnetic wave. The same approach was soon thereafter extended to optical forces and torques by Barton \textit{et. al} \cite{Barton1989a}, considering a focused beam model from Davis \cite{Davis1979,Barton1989b}. The Davis beam model depicts a paraxial Gaussian beam with higher order corrections.

From the initial development of GLMT, most applications required knowledge of the scattered field or beam shape coefficients, which are none other than the multipole expansion coefficients. Several strategies have been developed for the determination of the beam shape coefficients (BSCs) in the most efficient and/or accurate way. 

Apart from the several ways to determine the BSCs, at the heart of this development, the quest of the appropriate incident beam model was also highlighted. The suggested and employed beam models were not a solution to Maxwell's equation, and most of them were approximations that broke down for a highly focused beam.

Many researchers contributed to the understanding of optical forces by ways of theory and computational modeling \cite{Maia2000,Rohrbach2001,Lock2004a,Lock2004b,Nieminen2007,Chen2010,Nieminen2014,Gouesbet2017,Jones2015}. In this tutorial, we wish to present what we believe is a didactic approach towards an analytical solution, considering a physical and realistic trapping beam along with the exact determination of its beam shape coefficients. With the MST approach, an exact model for optical forces in a typical optical tweezers is obtained \cite{Neves2006OE}. Therefore, we merge the main contributions from some of the studies highlighted in this work, into what we believe is a robust and rigorous treatment for optical forces in optical tweezers. Similar approaches using BSC calculation via plane wave spectrum or for Gaussian beams independent of the radial coordinate, have been published \cite{Taylor2009,Salandrino2012,Siler2015,Lu2017}, but were unaware of the proposed method in Ref.~\cite{Neves2006OE}, 

A standard and modern notation for the equations was adopted throughout this work. The tutorial is divided into four parts. The first part introduces the EM fields along with the multipole coefficients. The second part explores how the BSCs are determined, and this is illustrated for the simple case of a plane wave. The third part presents the highly focused beam description along with the BSCs. Finally, the optical force through the Maxwell stress tensor is presented in terms of the BSCs. We hope to have addressed all the important points and developments described above.

\section{Modeling}

Hence, how do we start modeling the optical forces? Primarily, we rely on the knowledge of the exact electromagnetic (EM) fields, which travel as a wave with linear momentum. Because the trapped object is a microsphere, it seems reasonable to express the EM fields in terms of spherical harmonic functions, similarly to the standard textbook approach \cite{Jackson1999}. The encountered problem is essentially the elastic scattering of monochromatic coherent light by the homogeneous and isotropic sphere of radius ($a$).

\subsection{Description of the EM fields}

Let us describe all the required electromagnetic fields, incident (inc), scattered (sca), and internal (int) as time-harmonics in terms of spherical vector expansion, also known as multipole expansion, as our scattering particle has spherical symmetry \cite{Devaney1974}. The time dependence $e^{-i \omega t}$ is omitted for clarity. Our coordinate system has its origin at the center of our spherical scatterer, which scatters the incoming light. A similar approach with other notations is present in several textbooks \cite{Kerker1969,Bohren2008}.

\begin{equation}\label{eq:Einc}
\mathbf{E}_{\text{inc}}(\mathbf{r})=E_0
\sum_{l=1}^\infty\sum_{m=-l}^{l}
\left(
G_{lm}^{TE}
\mathbf{M}^{(1)}_{lm}(\mathbf{r})
+
G_{lm}^{TM}
\mathbf{N}^{(1)}_{lm}(\mathbf{r})
\right),
\end{equation}
\begin{equation}\label{eq:Hinc}
Z\mathbf{H}_{\text{inc}}(\mathbf{r})=E_0
\sum_{l=1}^\infty\sum_{m=-l}^{l}
\left(
G_{lm}^{TM}
\mathbf{M}^{(1)}_{lm}(\mathbf{r})
-G_{lm}^{TE}
\mathbf{N}^{(1)}_{lm}(\mathbf{r})
\right),
\end{equation}
\begin{equation}\label{eq:Escat}
\mathbf{E}_{\text{sca}}(\mathbf{r})=E_0
\sum_{l=1}^\infty\sum_{m=-l}^{l}
\left(
b_{lm} 
\mathbf{M}^{(3)}_{lm}(\mathbf{r})
+
a_{lm} 
\mathbf{N}^{(3)}_{lm}(\mathbf{r})
\right),
\end{equation}
\begin{equation}\label{eq:Hscat}
Z\mathbf{H}_{\text{sca}}(\mathbf{r})=E_0
\sum_{l=1}^\infty\sum_{m=-l}^{l}
\left(
a_{lm} 
\mathbf{M}^{(3)}_{lm}(\mathbf{r})
-b_{lm} 
\mathbf{N}^{(3)}_{lm}(\mathbf{r})
\right),
\end{equation}
\begin{equation}\label{eq:Eint}
\mathbf{E}_{\text{int}}(\mathbf{r})=E_0
\sum_{l=1}^\infty\sum_{m=-l}^{l}
\left(
d_{lm} 
\mathbf{M}^{(1)}_{lm}(\mathbf{r})
+
c_{lm} 
\mathbf{N}^{(1)}_{lm}(\mathbf{r})
\right),
\end{equation}
\begin{equation}\label{eq:Hint}
Z\mathbf{H}_{\text{int}}(\mathbf{r})=E_0
\sum_{l=1}^\infty\sum_{m=-l}^{l}
\left(
c_{lm} 
\mathbf{M}^{(1)}_{lm}(\mathbf{r})
-d_{lm} 
\mathbf{N}^{(1)}_{lm}(\mathbf{r})
\right),
\end{equation}

where $l$ and $m$ are the radial and azimuthal mode index, respectively. The electric field dimension constant $E_0 $ is used to make the expansion coefficients non--dimensional. The mode amplitudes, $a_{lm},b_{lm},c_{lm},d_{lm},G_{lm}^{TM},$ and $G_{lm}^{TE}$, are also called expansion coefficients or multipole coefficients, however,  the last pair is special, and we will refer to them as beam shape coefficients (BSCs). The vector spherical harmonics are, $k\mathbf{N}_{lm}(\mathbf{r}) =i\nabla\times\mathbf{M}_{lm}(\mathbf{r})$,
$\mathbf{M}_{lm}(\mathbf{r})=z_l(kr)\mathbf{X}_{lm}
(\mathbf{\hat{r}})$, and $\mathbf{X}(\mathbf{\hat{r}})=\mathbf{L}Y_{lm}
(\mathbf{\hat{r}})/\sqrt{l(l+1)}$ \cite{Hill1954}, where $z_l(kr)$ is the appropriate spherical Bessel function, $Y_{lm}(\mathbf{\hat{r}})$ are the scalar spherical harmonics, $Z=\sqrt{\mu /\varepsilon }$, $k=\omega \sqrt{\mu \varepsilon }$, and $\mathbf{L}=-i \mathbf{r}\times \nabla$ is the angular momentum operator in direct space. The above vector functions are solutions to the vector wave equation, first proposed by Hansen \cite{Hansen1935}, and any field can be expanded in these series given the appropriate coefficients. 

The spatial function $z^{(j)}_l(kr)$ is a basic solution to the spherical Bessel equation, which for $j=1,2,3,4$ corresponds to the ordinary spherical Bessel, spherical Neumann, and spherical Hankel functions of the first and second kind, respectively. The choice of the function depends on the particular spatial bounds, either for the fields inside or outside the scatterer. Because the fields must be regular at the origin, only $z_l^{(1)}(kr)=j_l(k_1r)$ can appear at the interior radial function, and also for the incident fields that exist in the origin, independent of the scatterer. While in the region outside the scatterer, we can write in terms of spherical Hankel functions of the outgoing type, based on the physical ground of the scattered wave, $z_l^{(3)}(kr)=h_l^{(1)}(k_1r)$ .

\subsection{Scattering coefficients}

To solve the electromagnetic boundary problem on the scatterer surface, the continuity of the tangential components of the electric and magnetic fields is required. The following equations need to be satisfied,
\begin{equation}
\left.\left[\mathbf{E}_{\text{inc}}(\mathbf{r})+\mathbf{E}_{\text{sca}}(\mathbf{r})-\mathbf{E}_{\text{int}}(\mathbf{r})\right]\times \mathbf{\hat{r}}\right|_{r=a}=0,
\end{equation}
\begin{equation}
\left.\left[\mathbf{H}_{\text{inc}}(\mathbf{r})+\mathbf{H}_{\text{sca}}(\mathbf{r})-\mathbf{H}_{\text{int}}(\mathbf{r})\right]\times \mathbf{\hat{r}}\right|_{r=a}=0.
\end{equation}
To solve each of these tangential components, let us introduce two new abbreviations,
\begin{equation}\label{eq:SizePar}
x=k_2a=\frac{2\pi n_2 a}{\lambda}, \qquad \qquad Mx=\frac{n_1}{n_2}k_2 a=k_1 a,
\end{equation}
where $x$ is the size parameter, $M$ is the relative refractive index, $n_1$ and $n_2$ are the refractive index of the particle and the surrounding medium, respectively. Note that the refractive index of the particle $n_1$ is allowed to be complex, to treat the effects of absorption such as the ones in metallic particles. For the electric TE modes, the solution is
\begin{equation}
\sum_{l=1}^\infty\sum_{m=-l}^{l}
\left(
G_{lm}^{TE} j_l(x)
+b_{lm} h_l^{(1)}(x)
-d_{lm} j_l(Mx)
\right)\mathbf{X}_{lm}(\mathbf{\hat{r}})\times \mathbf{\hat{r}}=0.
\end{equation}
This is valid for all angles, therefore
\begin{equation}
G_{lm}^{TE} j_l(x)+b_{lm} h_l^{(1)}(x)-d_{lm} j_l(Mx)=0.
\end{equation}
Analogously, for the magnetic TE modes, the condition valid for all angles is as follows
\begin{equation}
G_{lm}^{TM} j_l(x)+a_{lm} h_l^{(1)}(x)
-\frac{Z_1}{Z_2}c_{lm} j_l(Mx)=0.
\end{equation}
The terms proportional to $\mathbf{N}_{lm}(\mathbf{r})$ simplify to the following
\begin{equation}
\mathbf{N}_{lm}(\mathbf{r})\times \mathbf{\hat{r}}=\frac{i}{k}\left(\nabla\times(z_l(kr)\mathbf{X}_{lm}(\mathbf{\hat{r}}))\right)\times \mathbf{\hat{r}}=\frac{i}{kr}\frac{\partial (kr z_l(kr))}{\partial kr}.
\end{equation}

For the TM modes, we obtain
\begin{equation}
G_{lm}^{TM}
\frac{\partial (k_1r j_l(k_1r))}{\partial k_1r}
+a_{lm} 
\frac{\partial (k_1r h_l^{(1)})}{\partial k_1r}
-c_{lm} 
\frac{1}{M}\frac{\partial (k_2r j_l(k_2r))}{\partial k_2r}
=0.
\end{equation}

Moreover, for the magnetic equation, we have
\begin{equation}
G_{lm}^{TE} \frac{\partial (k_1r j_l(k_1r))}{\partial k_1r}
+ b_{lm} \frac{\partial (k_1r h_l^{(1)})}{\partial k_1r}
-\frac{Z_1}{Z_2}
 d_{lm} \frac{1}{M}\frac{\partial (k_2r j_l(k_2r))}{\partial k_2r}
=0.
\end{equation}

To further simplify the expressions above, we introduce the Riccati--Bessel functions,
\begin{equation}
\psi_l(y)=y\, j_l(y), \qquad \qquad \xi_l(y)=y\, h_l^{(1)}(y).
\end{equation}

Which leads to
\begin{equation}
G_{lm}^{TM} \psi_l'(x)
+a_{lm} \xi_l'(x)
-c_{lm} \frac{1}{M}\psi_l'(Mx)=0,
\end{equation}
and
\begin{equation}
G_{lm}^{TE} \psi_l'(x)
+b_{lm} \xi_l'(x)
-\frac{Z_1}{Z_2} \frac{1}{M} d_{lm} \psi_l'(Mx)=0.
\end{equation}

Considering the relations between the EM impedance and the waves at optical frequencies $\mu_1 \approx \mu_2$,
\begin{equation}
\frac{Z_1}{Z_2}\frac{1}{M}=\sqrt{\frac{\mu_1 \epsilon_2}{\epsilon_1 \mu_2}} \frac{n_1}{n_2}=\sqrt{\frac{\mu_1 \epsilon_2}{\epsilon_1 \mu_2}} \sqrt{\frac{\mu_1 \epsilon_1}{\mu_2 \epsilon_2}}\approx 1.
\end{equation}
The resulting four equations relating all multipole coefficients are:
\begin{align}
G_{lm}^{TE} \psi_l(x)+b_{lm}\xi_l(x)-\frac{1}{M}d_{lm}\psi_l(Mx)&=0, \label{eq:1}\\
G_{lm}^{TM} \psi_l(x)+a_{lm}\xi_l(x)-c_{lm}\psi_l(Mx)&=0, \label{eq:2}\\
G_{lm}^{TM} \psi_l'(x)+a_{lm} \xi_l'(x)-\frac{1}{M} c_{lm} \psi_l'(Mx)&=0, \label{eq:3}\\
G_{lm}^{TE} \psi_l'(x)+b_{lm} \xi_l'(x)-d_{lm} \psi_l'(Mx)&=0. \label{eq:4}
\end{align}
Because we are interested in the scattered fields, we can eliminate $c_{lm}$ and $d_{lm}$. First, we isolate $c_{lm}$ in Eq.(\ref{eq:2}) and substitute in Eq.(\ref{eq:3}).
\begin{equation}
a_{lm} =-G_{lm}^{TM} \left[\frac{M \psi_l'(x) \psi_l(Mx) -\psi_l(x) \psi_l'(Mx)}{M \xi_l'(x) \psi_l(Mx) -\xi_l(x)\psi_l'(Mx)}\right]=-G_{lm}^{TM} a_l.
\end{equation}
Secondly, we isolate $d_{lm}$ in Eq.(\ref{eq:4}) and substitute in Eq.(\ref{eq:1}).
\begin{equation}
b_{lm}=-G_{lm}^{TE} \left[\frac{M \psi_l(x) \psi_l'(Mx)-\psi_l'(x)\psi_l(Mx)}{M \xi_l(x) \psi_l'(Mx) -\xi_l'(x)\psi_l(Mx)}\right]=-G_{lm}^{TE} b_l,
\end{equation}

where the Mie scattering coefficients are thus defined as in Ref.\cite{Bohren2008},
\begin{align}
a_l&=\frac{M \psi_l'(x) \psi_l(Mx) -\psi_l(x) \psi_l'(Mx)}{M \xi_l'(x) \psi_l(Mx) -\xi_l(x)\psi_l'(Mx)},\\
b_l&=\frac{M \psi_l(x) \psi_l'(Mx)-\psi_l'(x)\psi_l(Mx)}{M \xi_l(x) \psi_l'(Mx) -\xi_l'(x)\psi_l(Mx)}.
\end{align}

A complete and efficient description of the scattered EM wave can be obtained if information about the scatterer and the incident field are available. The GLMT framework addresses the fields in all space (near and far fields), allowing for a complicated beam that satisfies Maxwell's equations. In our case, the scatterer is a homogeneous and isotropic dielectric microsphere, and it is completely characterized by the Mie scattering coefficients ($a_l$ and $b_l$), through the knowledge of the microsphere radius ($a$), the refractive index of the microsphere ($n_2$), the refractive index of the surrounding medium ($n_1$), and the wavelength of the incident wave ($\lambda$). For the incident beam, this information is present in the beam shaped coefficients ($G_{lm}^{TM}$ and $G_{lm}^{TE}$), which have been studied extensively by Gousbet, Grehan, and Lock \cite{Gouesbet1995,Gouesbet2017} (and references therein). The method by which we determine the beam shaped coefficients is presented in the following section.

\section{Beam shape coefficients}

The description of BSCs appears in the GLMT framework, which was initially pioneered when studying plasma velocity by laser Doppler velocimetry \cite{Gouesbet2017}. The results for the scattering of any beam are known once the beam shape coefficient is determined. In fact, GLMT can be applied to analytically solve scattering from other objects besides the homogeneous sphere, including the stratified sphere, spheroids, and infinite cylinders of any size. Moreover, any type of beam can be used, the solution not being restricted to Gaussian beams. The main problem in many studies and the subject of much debate is the calculation of the BSCs, mainly due to the non--cancellation of the radial spherical Bessel functions in both sides of the expansion. Primarily, they are defined as follows.

In Equations (\ref{eq:Einc}) to (\ref{eq:Hint}) we described all electromagnetic fields in terms of spherical vector expansion. In the previous section, we explicitly determined the complex scattering coefficients in terms of the beam shape coefficients (BSCs) $G_{lm}^{TM}$ and $G_{lm}^{TE}$. We can, therefore, express \textbf{any} incident EM field by determining these BSCs. Note that $\mathbf{M}_{lm}(\mathbf{r})$ is perpendicular to the radial vector, $\mathbf{\hat{r}}$, we can use this and the normalization relation of spherical harmonics;
\begin{equation}
\int Y_{lm}^* Y_{pq} d\Omega = \delta_{lp}\delta_{mq} ,
\label{eq:norm}
\end{equation}
where
\begin{equation}
Y_{lm}=\sqrt{\frac{2l+1}{4\pi}\frac{(l-m)!}{(l+m)!}} P_l^m(\cos(\theta))e^{i m \phi}.
\end{equation}
To determine the BSCs $G_{lm}^{TM}$ and $G_{lm}^{TE}$, we first multiply by the radial vector $\mathbf{r}$;
\begin{equation}
\mathbf{r}\cdot\mathbf{E}_{\text{inc}}(\mathbf{r})=E_0
\sum_{l=1}^\infty\sum_{m=-l}^{l}
G_{lm}^{TM}
\mathbf{r}\cdot\mathbf{N}^{(1)}_{lm}(\mathbf{r}).
\label{eq:rEinc}
\end{equation}
The last vector product in the equation above is determined by expanding in terms of spherical vector harmonics, and later in terms of spherical harmonics,
\begin{equation}
\mathbf{r}\cdot\mathbf{N}^{(1)}_{lm}(\mathbf{r})=-\frac{j_l(kr)}{k}\sqrt{l(l+1)}Y_{lm}.
\end{equation}
Therefore, Eq.(\ref{eq:rEinc}) becomes
\begin{equation}
\mathbf{r}\cdot\mathbf{E}_{\text{inc}}(\mathbf{r})=-\frac{E_0}{k}
\sum_{l=1}^\infty\sum_{m=-l}^{l}
G_{lm}^{TM}
j_l(kr)\sqrt{l(l+1)}Y_{lm}.
\end{equation}
To eliminate the double sum, we will multiply both sides by $Y_{pq}^*$ and integrate over the solid angle as follows
\begin{equation}
\begin{split}
&\int Y_{pq}^* \mathbf{r}\cdot\mathbf{E}_{\text{inc}}(\mathbf{r}) d\Omega \\
&=-\frac{E_0}{k}
\sum_{l=1}^\infty\sum_{m=-l}^{l}
G_{lm}^{TM}
j_l(kr)\sqrt{l(l+1)}\int Y_{pq}^* Y_{lm} d\Omega.
\end{split}
\end{equation}
Using the normalization property of Eq.(\ref{eq:norm}), we obtain
\begin{equation}
\int Y_{pq}^* \mathbf{r}\cdot\mathbf{E}_{\text{inc}}(\mathbf{r}) d\Omega=-\frac{E_0}{k}G_{pq}^{TM}
j_p(kr)\sqrt{p(p+1)},
\end{equation}
or
\begin{equation}\label{eq:DefGTM}
j_p(kr)G_{pq}^{TM}=-kr \frac{1}{E_0 \sqrt{p(p+1)}}\int Y_{pq}^* \mathbf{\hat{r}}\cdot\mathbf{E}_{\text{inc}}(\mathbf{r}) d\Omega.
\end{equation}
Analogous for the magnetic field, the corresponding BSC is given by
\begin{equation}\label{eq:DefGTE}
j_p(kr)G_{pq}^{TE}=kr \frac{Z}{E_0 \sqrt{p(p+1)}}\int Y_{pq}^* \mathbf{\hat{r}}\cdot\mathbf{H}_{\text{inc}}(\mathbf{r}) d\Omega.
\end{equation}

The Eqs.(\ref{eq:DefGTM}) and (\ref{eq:DefGTE}) above define the beam shape coefficients for the incident trapping beam from the GLMT framework. The right--hand side of these equations should contain a spherical Bessel function from the integration to cancel with the one on the left--hand side. The major shortcoming of the other methods is that this radial dependent term has never been explicitly determined to produce BSC independent of the radial coordinates.

The radial component of the electric and magnetic fields of the incident beam needs to be determined. Once the exact EM is known, or described an analytical expression, we are able to recover an exact BSCs. However, if the incident EM fields is only approximated, we no longer retain an exact BSC. This is the case for the Davis beam, which results in a BSCs that depends on the radial coordinate ($r$). By definition, the expansion coefficient should not depend on the radial coordinate, hence Barton suggested to determine the BSCs at the particle surface ($r=a$) \cite{Barton1989a}. However, what would happen if $ka$ is a root of the spherical Bessel function ($j_l(ka)$)? Note that there is an infinite sum in the order $l$ and infinite roots for each order.

\subsection{BSC for a plane wave}

The simplest incident field is a plane wave. Here, we determine the BSCs of the plane wave by direct integration. Consider an incident coherent monochromatic plane wave propagating in non-absorbing homogeneous media in the z--direction, which is linearly polarized in the x--axis,
\begin{equation}
\mathbf{E}_{\text{inc}}= E_0 e^{i kz} \mathbf{\hat{x}}.
\end{equation}
In spherical coordinates, the radial component is
\begin{equation}\label{eq:PW}
\mathbf{\hat{r}}\cdot\mathbf{E}_{\text{inc}}(\mathbf{r})=E_0 e^{i kr \cos\theta} \sin\theta\cos\phi.
\end{equation}

We can now calculate the plane wave BSCs from the definition of Eq.(\ref{eq:DefGTM}),
\begin{equation}
\begin{split}
G_{lm}^{TM}&=-\frac{kr}{j_l(kr)}\frac{1}{\sqrt{l(l+1)}}\int Y_{lm}^* e^{i kr \cos\theta} \sin\theta\cos\phi d\Omega \\
&=-\frac{kr}{j_l(kr)}\frac{1}{\sqrt{l(l+1)}}\sqrt{\frac{2l+1}{4\pi}\frac{(l-m)!}{(l+m)!}}\\
&\int_{\theta=0}^{\pi} P_l^m(\cos\theta) e^{i kr \cos\theta}  \sin^2\theta d\theta  \int_{\phi=0}^{2\pi} e^{-i m \phi} \cos\phi d\phi .
\end{split}
\end{equation}
The solution for the integral in $\phi$ is,
\begin{equation}
\int_{\phi=0}^{2\pi} e^{-i m \phi} \cos\phi d\phi =\pi (\delta_{m,+1}+\delta_{m,-1}).
\end{equation}
Therefore, for plane waves only BSCs with $m=\pm 1$ remains. Let us determine the BSC for $m=+1$.
\begin{equation}
G_{l,+1}^{TM}=-\frac{kr}{j_l(kr)}\frac{\pi}{l(l+1)}\sqrt{\frac{2l+1}{4\pi}}\int_{\theta=0}^{\pi} P_l^1(\cos\theta) e^{i kr \cos\theta}  \sin^2\theta d\theta.
\label{eq:GTM1}
\end{equation}

The $\theta$--integral can be solved using the integral representation of the spherical Bessel function,

\begin{equation}
\int_{\theta=0}^{\pi} \sin^{2}\theta e^{i kr \cos\theta} P_l^1(\cos\theta) d\theta = 2 i^{l+1} l(l+1) \frac{j_l(kr)}{kr}.
\end{equation}
Eq.(\ref{eq:GTM1}) thus becomes
\begin{equation}
G_{l,+1}^{TM}=-\sqrt{\pi(2l+1)} i^{l+1}.
\end{equation}
For $m=-1$, we have
\begin{equation}
G_{l,-1}^{TM}=\sqrt{\pi(2l+1)} i^{l+1}.
\end{equation}
Similarly, for the other BSC,
\begin{equation}
\begin{split}
G_{lm}^{TE}&=\frac{kr}{j_l(kr)}\frac{1}{\sqrt{l(l+1)}}\int Y_{pq}^* e^{i kr \cos\theta} \sin\theta\sin\phi d\Omega \\
&=\frac{kr}{j_l(kr)}\frac{1}{\sqrt{l(l+1)}}\sqrt{\frac{2l+1}{4\pi}\frac{(l-m)!}{(l+m)!}}\\
&\int_{\theta=0}^{\pi} P_l^m(\cos\theta) e^{i kr \cos\theta}  \sin^2\theta d\theta  \int_{\phi=0}^{2\pi} e^{-i m \phi} \sin\phi d\phi .
\end{split}
\end{equation}
The $\phi$ integral now becomes
\begin{equation}
\int_{\phi=0}^{2\pi} e^{-i m \phi} \sin\phi d\phi =-i\pi (\delta_{m,+1}-\delta_{m,-1}).
\end{equation}
For $m=\pm1$,
\begin{equation}
G_{l,\pm1}^{TE}=-i \sqrt{\pi(2l+1)} i^{l+1}.
\end{equation}

We can verify that these are the true BSC, by inserting them into the radial component of the incident electric field Eq.(\ref{eq:Einc}).

\begin{equation}\label{eq:PWExpansion}
\begin{split}
&\mathbf{\hat{r}}\cdot\mathbf{E}_{\text{inc}}(\mathbf{r})=E_0
\sum_{l=1}^\infty\sqrt{\pi(2l+1)} i^{l+1} \mathbf{\hat{r}}\cdot\left(
-\mathbf{N}_{l,+1}(\mathbf{r})
+\mathbf{N}_{l,-1}(\mathbf{r})
\right)\\
&=E_0 \sum_{l=1}^\infty \frac{j_l(kr)}{kr}\sqrt{l(l+1)}\sqrt{\pi(2l+1)} i^{l+1}\left(-Y_{l,+1}+Y_{l,-1}\right)\\
&=-i \frac{E_0}{kr} \cos\phi\sin\theta\frac{d}{d\,(\cos\theta)}\sum_{l=1}^\infty j_l(kr) i^l\left(2l+1\right)P_l(\cos\theta)\\
&=E_0 e^{i kr \cos\theta} \sin\theta\cos\phi.
\end{split}
\end{equation}

The last summation in Eq.(\ref{eq:PWExpansion}) is a well--known relation, known as the plane wave expansion or Rayleigh equation, which describes the plane wave as a sum of spherical waves. As a result, we recover the expected electric field of Eq.(\ref{eq:PW}). Having explained how to obtain the plane wave BSC, let us turn to the beam of interest for optical tweezers, namely the highly focused Gaussian beam.

\section{Highly focused Gaussian beam}

Another great challenge for a proper beam description is using an expression for the EM field that satisfies exactly Maxwell's Equation. The beam representation must represent a solution of the vector Helmholtz equation, and the paraxial Gaussian beam is not a solution. The use of these pseudo-fields \cite{Nieminen2003}, or non-Maxwellian beams, would result in an approximate beam description leading to a pair of BSCs (Eqs.(\ref{eq:DefGTM}) and (\ref{eq:DefGTE})) that depend on the radial coordinate \cite{Neves2006OL}. 

Early studies modeled a focused Gaussian beam as a paraxial beam perturbed up to the fifth-order through a series expansion in terms of a $s-$parameter ($s=1/(kw_0)$, where $w_0$ is the beam waist at the focus) \cite{Barton1989b}. This Davis-Barton fifth-order approximation still does not solve Maxwell's equation exactly and fails to converge to reproduced experimental results under highly focused conditions ($s\rightarrow1$) as is the case in optical trapping \cite{Wright1993,Novotny1998}. The main limitation of beam models is that they do not satisfy exactly Maxwell's equation, and do not represent a good approximation of real beam when the spot size is too small. Most paraxial beams used in focusing are also non--physical because they do not exhibit a change in spot size due to polarization alone \cite{Dorn2003}. Moreover, the diffraction effects due to the truncated Gaussian beam by the finite objective lens and the polarization mixing by the high numerical aperture (NA) objective are not correctly represented near the focus of a high-NA objective \cite{Rohrbach2001,Ganic2004}.

The suggested most efficient way to overcome this issue of an accurate beam model for optical force prediction was to treat the high-NA focusing as a vector diffraction problem \cite{Wright1993}. 

At high-NA focusing, the paraxial scalar theory does not take into account the vector nature of the beam, it does not justify small-angle approximations, and it does not consider the apodization effects of optical systems \cite{McCutchen1964,Sheppard1987}. This has been thoroughly tested, proven and compared in numerous experiments, such as the 4Pi confocal fluorescence microscope \cite{Hell1992}, which led to the development of super--resolved fluorescence microscopy. A tutorial for diffraction calculation in optical systems has recently been published \cite{Kim2018}, discussing diffraction integrals and high NA focusing.

The solution for the highly focused beam that represents the vector property of the incident beam is the vectorial Debye–Wolf diffraction integral method, proposed by Richards and Wolf \cite{Richards1959}. This model describes the paraxial Gaussian beam being focused by a highly convergent aplanatic lens as a sum of plane waves arriving from all the angles spanned by the lens´s numerical aperture as an angular spectrum of plane waves \cite{Novotny2012}. Recently, it was highlighted that this approach to vector diffraction dates to a much earlier publication, in 1920 by Ignatovsky \cite{Peatross2017,Ignatowsky1919}.

\subsection{Angular Spectrum representation of focused Gaussian beams}

The angular spectrum representation depicts a mathematical instrument to calculate the electromagnetic fields near a highly focused beam (by a lens of focal length $f$). It comprises the decomposition of the EM field into an infinite series expansion in terms of Fourier components. With respect to optical tweezers, we are concerned with the propagating part of the angular spectrum, which can be solved with the method of stationary phase \cite{Mandel1995}. 

The resulting electric field in the vicinity of the focus ($\mathbf{E_{inc}}(\rho,\varphi,z)$), as lens focus is chosen as the origin of the coordinate system, with respect to the field on a reference sphere ($\mathbf{E_{\infty}}(\theta,\phi)$) is given in cylindrical coordinates as follows:
\begin{equation}\label{eq:ASR}
\begin{split}
&\mathbf{E}_{\text{inc}}(\rho,\varphi,z)=\frac{i k f e^{ikf}}{2\pi}\\
&\int_{0}^{\theta_{\text{NA}}} \int_{0}^{2\pi} \mathbf{E_{\infty}}(\theta,\phi) e^{ikz\cos\theta} e^{ik\rho\sin\theta\cos(\phi-\varphi)}\sin\theta\,d\theta\,d\phi,
\end{split}
\end{equation}
where $\theta_\text{NA}=\sin^{-1}(\text{NA}/n_a)$ is the maximum acceptance angle given by the NA of the objective lens. The field on the reference sphere ($\mathbf{E_{\infty}}$) is itself determined from the paraxial Gaussian $TEM_{00}$ beam as
\begin{equation}\label{eq:Aplanatic}
\mathbf{E_{\infty}}= \sqrt{\frac{n_b}{n_a}\cos\theta}\left(\left[ \mathbf{E}_{\text{par}} \cdot \boldsymbol{\hat{\varphi}}\right] \boldsymbol{\hat{\phi}}+\left[ \mathbf{E}_{\text{par}} \cdot \boldsymbol{\hat{\rho}}\right]\boldsymbol{\hat{\theta}}\right),
\end{equation}

where $n_b$ and $n_a$ refer to the refractive index of the medium before and after the focusing lens, respectively.  The electric field in the focal region, Eq.(\ref{eq:ASR}), is described in cylindrical coordinates and completely satisfies Maxwell's equation. The angular spectrum of a Gaussian beam requires a corresponding Gaussian spectrum that includes evanescent components. These non-homogeneous components decay exponentially with distance from the reference sphere. In optical tweezers, the focal region would be far-field from the reference sphere. Therefore, this method allows for a more rigorous description of the focused field in comparison to the previous ones \cite{Novotny2012}.

To determine the BSCs, we need to convert to spherical coordinates. We would also like to displace this beam from its origin (the focal point) to an arbitrary point in space ($\rho_0$,$\varphi_0$,$z_0$). This is achieved with the following substitutions:
\begin{align}
\rho\cos\varphi&=r\sin\theta\cos\phi-\rho_0\cos\phi_0,\\
\rho\sin\varphi&=r\sin\theta\sin\phi-\rho_0\sin\phi_0,\\
z&=r\cos\theta-z_0.
\end{align}

\begin{figure}[h!]
\centering\includegraphics[width=0.9\textwidth]{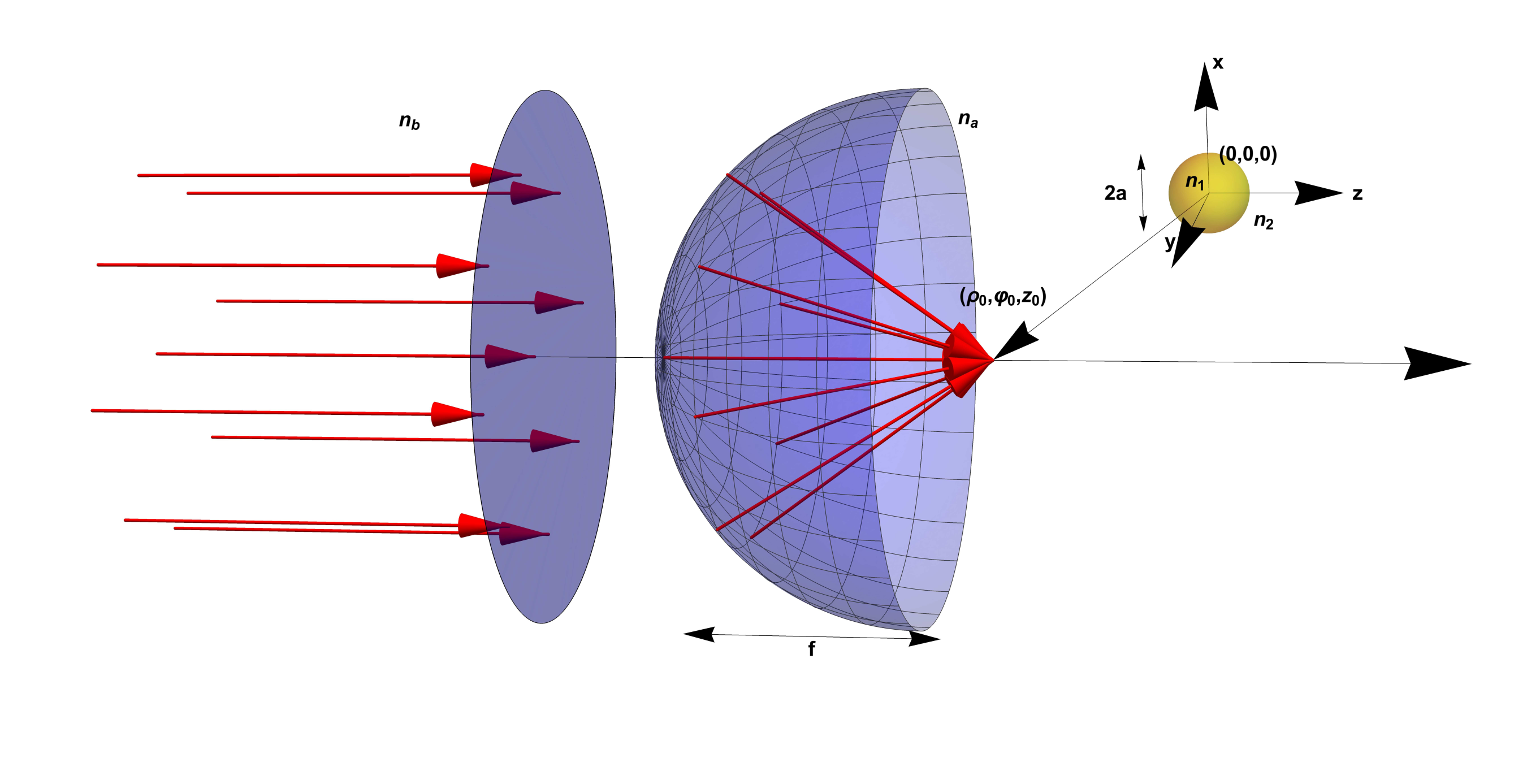}
\caption{Illustration of the variables from the incoming beam (red rays), mapped onto a reference sphere and later focused (converging red rays). The focus is shifted from the origin of coordinates located at the center of the spherical scatterer.}
\end{figure}

The purpose of shifting the EM field with respect to the sphere origin provides a later advantage in describing the scattering, or the corresponding optical forces, when the focused beam is displaced with respect to the sphere center. In general particle scattering theory, such as the transition matrix (T-matrix) is formulated in the coordinate system located at the particle center \cite{Mackowski1996,Borghese2007}. The beam (and it's BSCs) is described with respect to another origin with the use of the translational--addition theorem \cite{Cruzan1962,Doicu1997,Moine2005}, which adds another double sum operation that needs to be truncated appropriately. In our case, this is avoided by first generalizing the focused beam description (and its corresponding BSCs) with respect to any arbitrary location.

We now modify the spherical angular coordinates of the focusing system output pupil to avoid confusion with the BSCs integration variables, from ($\theta$,$\phi$) to ($\alpha$,$\beta$). The focused field becomes

\begin{equation}
\begin{split}
&\mathbf{E}_{\text{inc}}(\rho,\varphi,z)=\frac{i k f e^{ikf}}{2\pi}\int_{0}^{\alpha_\text{NA}}\,d\alpha \sin\alpha e^{ikr\cos\alpha\cos\theta} e^{-ikz_0\cos\alpha}\\
& \int_{0}^{2\pi}\,d\beta \mathbf{E_{\infty}}(\alpha,\beta) e^{ikr\sin\alpha\sin\theta\cos(\beta-\phi)} e^{-ik\rho_0\sin\alpha\cos(\beta-\phi_0)} .
\end{split}
\end{equation}

The amplitude of the incoming field, before focusing, has a Gaussian distribution and general polarization:

\begin{equation}
\mathbf{E}_{\text{par}}= E_0 e^{- f^2 \sin^2 \alpha/\omega^2} \left(p_x \mathbf{\hat{x}}+p_y\mathbf{\hat{y}}\right),
\end{equation}

where $\omega$ is the beam waist at the objective aperture, and ($p_x$,$p_y$) are the complex polarization parameters describing the state of polarization of the beam before entering an aplanatic objective lens. For example, linear polarization in the x-axis corresponds to (1,0). Consequently, the magnetic field is given as follows:

\begin{equation}
Z \mathbf{H}_{\text{par}}=\mathbf{\hat{z}}\times\mathbf{E}_{\text{par}}=E_0 e^{- f^2 \sin^2 \alpha/\omega^2}\left(p_x \mathbf{\hat{y}}-p_y\mathbf{\hat{x}}\right).
\end{equation}

That is, to switch to the BSC which depends on the magnetic field we have to change $p_x\rightarrow-p_y$ and $p_y\rightarrow p_x$.
Hence, we can determine Eq.(\ref{eq:Aplanatic}) with the above incident field, changing to the new $\alpha$ and $\beta$ angle coordinates. For the BSC, we only require the radial component of the incident field, which implies the radial component of the field at the reference sphere,

\begin{equation}
\mathbf{\hat{r}}\cdot\mathbf{E}_{\infty}=E_{x,\infty}\sin\theta\cos\phi+E_{y,\infty}\sin\theta\sin\phi+E_{z,\infty}\cos\theta.
\end{equation}

Thus, the field at the reference sphere and the radial component at the focus are,

\begin{equation}
\begin{split}
\mathbf{\hat{r}}\cdot\mathbf{E}_{\infty}(\alpha,\beta)&= \sqrt{\frac{n_b}{n_a}\cos\alpha} \, E_0 e^{- f^2 \sin^2 \alpha/\omega^2}\left(\right.\\
&\left[-\sin\beta\left[-p_x\sin\beta+p_y\cos\beta\right]+\cos\beta\cos\alpha\left[ p_x\cos\beta+p_y\sin\beta\right]\right]\sin\theta\cos\phi\\
&+\left[\cos\beta\left[-p_x\sin\beta+p_y\cos\beta\right]+\sin\beta\cos\alpha\left[ p_x\cos\beta+p_y\sin\beta\right]\right]\sin\theta\sin\phi\\
&\left.-\sin\alpha\left[ p_x\cos\beta+p_y\sin\beta\right]\cos\theta
\right),
\end{split}
\end{equation}

\begin{equation}\label{eq:FocusedField}
\begin{split}
\mathbf{\hat{r}}\cdot\mathbf{E}_{\text{inc}}(\rho,\varphi,z)&=\frac{i k f e^{ikf} E_0}{2\pi}\sqrt{\frac{n_b}{n_a}} \int_{0}^{\alpha_\text{NA}} d\alpha\, \sin\alpha \sqrt{\cos\alpha} e^{- f^2 \sin^2 \alpha/\omega^2} e^{-ikz_0\cos\alpha} e^{ikr\cos\alpha\cos\theta} \\
&\int_{0}^{2\pi} d\beta \,e^{ikr\sin\alpha\sin\theta\cos(\beta-\phi)} e^{-ik\rho_0\sin\alpha\cos(\beta-\phi_0)} \\
& \left\{\left[-\sin\beta\left(-p_x\sin\beta+p_y\cos\beta\right)+\cos\beta\cos\alpha\left( p_x\cos\beta+p_y\sin\beta\right)\right]\sin\theta\cos\phi\right.\\
&+\left[\cos\beta\left(-p_x\sin\beta+p_y\cos\beta\right)+\sin\beta\cos\alpha\left( p_x\cos\beta+p_y\sin\beta\right)\right]\sin\theta\sin\phi\\
&\left.-\sin\alpha \left( p_x\cos\beta+p_y\sin\beta \right) \cos\theta\right\}.
\end{split}
\end{equation}

These double integrals are addressed in the following section.

\subsection{Beam shape coefficients of focused Gaussian beam}

We can now determine the BSCs using the definition in Eq.(\ref{eq:DefGTM}), with the radial component of the incident field described in Eq.(\ref{eq:FocusedField}).

\begin{equation}
\begin{split}
G_{lm}^{TM}&=-\frac{kr}{j_l(kr)}\frac{1}{E_0 \sqrt{l(l+1)}} \sqrt{\frac{2l+1}{4\pi}\frac{(l-m)!}{(l+m)!}} \int_{0}^{\pi} \,d\theta P_l^m(\cos\theta) \sin\theta \int_{0}^{2\pi} \,d\phi e^{-i m \phi} \, \mathbf{\hat{r}}\cdot\mathbf{E}_{\text{inc}}(\mathbf{r})\\
&=-\frac{kr}{j_l(kr)}\frac{i k f e^{ikf}}{2\pi}\sqrt{\frac{n_b}{n_a}}\sqrt{\frac{2l+1}{4\pi l(l+1)}\frac{(l-m)!}{(l+m)!}} 
\int_{0}^{\alpha_\text{NA}} d\alpha \, \sin\alpha \sqrt{\cos\alpha} e^{- f^2 \sin^2 \alpha/\omega^2} e^{-ikz_0\cos\alpha} \\
&\int_{0}^{2\pi} d\beta \,e^{-ik\rho_0\sin\alpha\cos(\beta-\phi_0)} \int_{0}^{\pi} d\theta \, P_l^m(\cos\theta) \sin\theta e^{ikr\cos\alpha\cos\theta} \int_{0}^{2\pi} d\phi \, e^{-i m \phi} e^{ikr\sin\alpha\sin\theta\cos(\beta-\phi)}\\
&\left\{\left[-\sin\beta\left(-p_x\sin\beta+p_y\cos\beta\right)+\cos\beta\cos\alpha\left(p_x\cos\beta+p_y\sin\beta\right)\right]\sin\theta\cos\phi\right.\\
&+\left[\cos\beta\left(-p_x\sin\beta+p_y\cos\beta\right)+\sin\beta\cos\alpha\left( p_x\cos\beta+p_y\sin\beta\right)\right]\sin\theta\sin\phi\\
&\left.-\sin\alpha\left[ p_x\cos\beta+p_y\sin\beta\right]\cos\theta\right\}.
\end{split}
\end{equation}

We proceed by solving three of the four integrals above. The first integral to solve is that of $\phi$, where we use the following relation,

\begin{equation}
\begin{split}
&\int_{0}^{2\pi} \,d\phi e^{-i m \phi} e^{ix\cos(\beta-\phi)}\begin{pmatrix}\cos\phi\\\sin\phi\\1\end{pmatrix}=\\
&=-2\pi i^m e^{-im\beta}\begin{pmatrix}(-\sin\beta\frac{m J_{m}(x)}{x}+i\cos\beta\frac{d\,J_{m}(x)}{d x})\\(\cos\beta\frac{m J_{m}(x)}{x}+i\sin\beta\frac{d\,J_{m}(x)}{d x})\\-J_{m}(x)\end{pmatrix},
\end{split}
\end{equation}

where we abbreviated $y=kr\sin\alpha\sin\theta$. This solves the first of four integrals, and we will further introduce some dummy variables to shorten the algebra.

\begin{equation}
\begin{split}
A&=i^{m+1} k f e^{ikf} \frac{kr}{j_l(kr)} \sqrt{\frac{n_b}{n_a}}\sqrt{\frac{2l+1}{4\pi l(l+1)}\frac{(l-m)!}{(l+m)!}},\\
B&=\left[-p_x\sin\beta+p_y\cos\beta\right],\\
C&=\left[ p_x\cos\beta+p_y\sin\beta\right].
\end{split}
\end{equation}

which after rearranging term proportional to $\cos\alpha$ becomes

\begin{equation}
\begin{split}
&G_{lm}^{TM}=A \int_{0}^{\alpha_\text{NA}} \,d\alpha \sin\alpha \sqrt{\cos\alpha} e^{- f^2 \sin^2 \alpha/\omega^2} e^{-ikz_0\cos\alpha} \\
&\int_{0}^{2\pi} \,d\beta e^{-ik\rho_0\sin\alpha\cos(\beta-\phi_0)} e^{-im\beta}
\int_{0}^{\pi} \,d\theta P_l^m(\cos\theta) \sin\theta e^{ikr\cos\alpha\cos\theta} \\
&\left(B\sin\theta\frac{m J_{m}(y)}{y}+iC\cos\alpha\sin\theta\frac{d\,J_{m}(y)}{d y}+C\sin\alpha\cos\theta J_{m}(y)\right).
\end{split}
\end{equation}

Both $C$ terms can be combined within the $\theta$ integral, considering it as a result of the derivative with respect to $\alpha$,

\begin{equation}
\begin{split}
G_{lm}^{TM}&=A \int_{0}^{\alpha_\text{NA}} \,d\alpha \sin\alpha \sqrt{\cos\alpha} e^{- f^2 \sin^2 \alpha/\omega^2} e^{-ikz_0\cos\alpha} \\
&\frac{1}{kr}\int_{0}^{2\pi} \,d\beta e^{-ik\rho_0\sin\alpha\cos(\beta-\phi_0)} e^{-iq\beta}
\left[\frac{mB}{\sin\alpha}+iC\frac{d }{d \alpha}\right]\\
&\int_{0}^{\pi} \,d\theta P_l^m(\cos\theta) \sin\theta e^{ikr\cos\alpha\cos\theta} J_{m}(kr\sin\alpha\sin\theta).
\end{split}
\end{equation}

The result for the $\theta$--integral is given in Ref.~\cite{Neves2006JPA} as follows,
\begin{equation}\label{eq:ThetaIntegral}
\begin{split}
&\int_{0}^{\pi} \,d\theta P_l^m(\cos\theta) \sin\theta e^{ikr\cos\alpha\cos\theta} J_{m}(kr\sin\alpha\sin\theta)=\\
&=2 i^{l-m} P_l^m(\cos\alpha) j_l(kr).
\end{split}
\end{equation}

This reduces the BSC to only two integrals,
\begin{equation}
\begin{split}
G_{lm}^{TM}&=2 i^{l-m} \frac{j_l(kr)}{kr} A \int_{0}^{\alpha_\text{NA}} \,d\alpha \sin\alpha \sqrt{\cos\alpha} e^{- f^2 \sin^2 \alpha/\omega^2} e^{-ikz_0\cos\alpha}\\
&\int_{0}^{2\pi} \,d\beta e^{-ik\rho_0\sin\alpha\cos(\beta-\phi_0)} e^{-im\beta}\left[\frac{m B P_l^m(\cos\alpha) }{\sin\alpha}+iC\frac{d P_l^m(\cos\alpha)}{d \alpha}\right].
\end{split}
\end{equation}

Note that this last $\theta$-integral has actually cancelled the radial dependence from the spherical Bessel function defined in the BSC expression of Eq.(\ref{eq:DefGTM}). The analytical result for this integral is the key to avoid most numerical approximation in literature. This integral has been pointed out as a special case of a Gegenbauer integral \cite{Koumandos2007}, even though it is not generally well known and not easily available in integral tables \cite{Cregg2007}. This result has been useful in several applications other than light scattering, including acoustics \cite{Koyama2016,Gong2017}, ionization rates for orbitals \cite{Barth2011}, the Casimir interaction \cite{Messina2015}, atom tunneling \cite{Braun2008}, field-theory \cite{Polychronakos2011}, and cosmology \cite{Pillado2010}.

This analytical calculation was not derived for many years, and other strategies have been employed to solve the problem of determining BSCs, such as the use of van de Hulst’s localization principle, building a localized model of the beam. For an incident Gaussian beam on-axis to a sphere, described by the Davis model, the transverse and angular (polar) coordinates are mapped to specific values within the localized approximation as, $kr\rightarrow n+1/2$ and $\theta\rightarrow \pi/2$, respectively, thereby no longer depending on the radial variable, and avoiding the integral over the polar component \cite{Ren1998,Gouesbet1999,Lock2013}. A historical overview of the different methods for determining the BSC can be found in Gouesbet's book \cite{Gouesbet2017}.

To proceed with the determination of BSCs, we integrate (with respect to $\beta$), as we need to eliminate our dummy indexes that are functions of $\beta$. We introduce two angular functions related to the associated Legendre function, which are common in light scattering \cite{Kerker1969},

\begin{equation}
\pi_l^m(\alpha)=\frac{m P_l^m(\cos\alpha)}{\sin\alpha}, \qquad \qquad \tau_l^m(\alpha)=\frac{d P_l^m(\cos\alpha)}{d \alpha}.
\end{equation}

After some algebra we obtain,

\begin{equation}
\begin{split}
G_{lm}^{TM}&=i^{l+1} k f e^{ikf}\sqrt{\frac{n_b}{n_a}}\sqrt{\frac{2l+1}{\pi l(l+1)}\frac{(l-m)!}{(l+m)!}} \\
&\int_{0}^{\alpha_\text{NA}} \,d\alpha \sin\alpha \sqrt{\cos\alpha} e^{- f^2 \sin^2 \alpha/\omega^2} e^{-ikz_0\cos\alpha}\\
&\int_{0}^{2\pi} \,d\beta e^{-ik\rho_0\sin\alpha\cos(\beta-\phi_0)} e^{-im\beta}\\
&\left[\left(-p_x \pi_l^m(\alpha)+i p_y \tau_l^m(\alpha)\right)\sin\beta+\left(p_y \pi_l^m(\alpha)+i p_x \tau_l^m(\alpha)\right)\cos\beta\right].
\end{split}
\end{equation}

The integral (with respect to $\beta$) is equal to,

\begin{equation}
\begin{split}
&\int_{0}^{2\pi} \,d\beta e^{-ix\cos(\beta-\phi_0)} e^{-im\beta} \begin{pmatrix}
\cos\beta\\\sin\beta\end{pmatrix}=\\
&2\pi i^{-m} e^{-im\phi_0} \begin{pmatrix}
-\sin\phi_0 \frac{m J_m(x)}{x}+i\cos\phi_0 \frac{d J_m(x)}{d x}\\
\cos\phi_0 \frac{m J_m(x)}{x}+i\sin\phi_0 \frac{d J_m(x)}{d x}
\end{pmatrix},
\end{split}
\end{equation}

where $x=k\rho_0\sin\alpha$. Solving this last analytical integral, we obtain the BSC for the highly focused incident beam.

\begin{equation}
\begin{split}
G_{lm}^{TM}&=i^{l-m+1} k f e^{ikf}e^{-im\phi_0}\sqrt{\frac{n_b}{n_a}}\sqrt{\frac{4\pi (2l+1)}{l(l+1)}\frac{(l-m)!}{(l+m)!}} \\
&\int_{0}^{\alpha_\text{NA}} \,d\alpha \sin\alpha \sqrt{\cos\alpha} e^{- f^2 \sin^2 \alpha/\omega^2} e^{-ikz_0\cos\alpha}\\
&\left\{\left[-p_x \pi_l^m(\alpha)+i p_y \tau_l^m(\alpha)\right]\left[\cos\phi_0 \frac{m J_m(x)}{x}+i\sin\phi_0 \frac{d J_m(x)}{d x}\right]\right.\\
&\left.+\left[p_y \pi_l^m(\alpha)+i p_x \tau_l^m(\alpha)\right]\left[-\sin\phi_0 \frac{m J_m(x)}{x}+i\cos\phi_0 \frac{d J_m(x)}{d x}\right]\right].
\end{split}
\end{equation}

The analogous TE BSC,

\begin{equation}
\begin{split}
G_{lm}^{TE}&=-i^{l-m+1} k f e^{ikf}e^{-im\phi_0}\sqrt{\frac{n_b}{n_a}}\sqrt{\frac{4\pi (2l+1)}{l(l+1)}\frac{(l-m)!}{(l+m)!}} \\
&\int_{0}^{\alpha_\text{NA}} \,d\alpha \sin\alpha \sqrt{\cos\alpha} e^{- f^2 \sin^2 \alpha/\omega^2} e^{-ikz_0\cos\alpha}\\
&\left\{\left[p_y \pi_l^m(\alpha)+i p_x \tau_l^m(\alpha)\right]\left[\cos\phi_0 \frac{m J_m(x)}{x}+i\sin\phi_0 \frac{d J_m(x)}{d x}\right]\right.\\
&\left.+\left[p_x \pi_l^m(\alpha)-i p_y \tau_l^m(\alpha)\right]\left[-\sin\phi_0 \frac{m J_m(x)}{x}+i\cos\phi_0 \frac{d J_m(x)}{d x}\right]\right].
\end{split}
\end{equation}

This numerical integration is quite fast once the integrating functions are known. For the calculation of optical forces with the beam positioned in space ($\rho_0$,$\phi_0$,$z_0$) with respect to the sphere center, only the BSCs need to be recalculated, which is an advantage with respect to other methods that do not share this feature. These BSCs can be expressed in a more compact form, uncoupling the azimuthal coordinate ($\phi_0$) and polarization ($p_x$,$p_y$) \cite{Neves2007},

\begin{equation}\label{eq:GTMTE}
\begin{split}
&G_{lm}^{TM,TE}=-i^{l-m+1} k f e^{ikf}e^{-im\phi_0}\sqrt{\frac{n_b}{n_a}}\sqrt{\frac{4\pi (2l+1)}{l(l+1)}\frac{(l-m)!}{(l+m)!}} \int_{0}^{\alpha_\text{NA}} \,d\alpha \sin\alpha \sqrt{\cos\alpha} e^{- f^2 \sin^2 \alpha/\omega^2} e^{-ikz_0\cos\alpha}\\
&\left(
\begin{array}{cc}
 \left[\frac{m J_m(k\rho_0\sin\alpha)}{k\rho_0\sin\alpha} \pi_l^m(\alpha)+ J'_m(k\rho_0\sin\alpha) \tau_l^m(\alpha)\right], & -i\left[\frac{m J_m(k\rho_0\sin\alpha)}{k\rho_0\sin\alpha} \tau_l^m(\alpha)+ J'_m(k\rho_0\sin\alpha) \pi_l^m(\alpha)\right]
\end{array}
\right)\\
&.\left(
\begin{array}{cc}
 \cos\phi_0 & \sin\phi_0 \\
 -\sin\phi_0 & \cos\phi_0
\end{array}
\right).\left(
\begin{array}{c}
p_x,p_y \\
p_y,-p_x
\end{array}
\right).
\end{split}
\end{equation}

The BSCs $G_{lm}^{TM,TE}$ corresponds to the transverse magnetic (TM) and transverse electric (TE) multipoles, which correspond to the left and --right polarizations of the last vector in Eq.(\ref{eq:GTMTE}), respectively, whose last lines involve a (1$\times$2)(2$\times$2)(2$\times$1) matrix operation. 
The efficiency of this calculation can be higher by exploiting the symmetry of some special cases, as highlighted in the following sections.

\subsection{Case study: coincident microsphere and beam axes}

This case is a common condition under trapping equilibrium, resulting in the microsphere axis coincident with the beam axis. To determine these special BSCs, we make, $x\rightarrow0$, which leads to the following conditions;
\begin{equation}
\begin{split}
\lim_{x\to 0} \frac{d J_m(x)}{d x} &= \frac{1}{2}(\delta_{m,+1}-\delta_{m,-1}), \\
\lim_{x\to 0} \frac{m J_m(x)}{x} &= \frac{1}{2} (\delta_{m,+1}+\delta_{m,-1}). 
\end{split}
\end{equation}

For $m=\pm 1$ we have

\begin{equation}
\begin{split}
&G_{l,\pm 1}^{TM}=\frac{i^l (k f e^{ikf})}{l(l+1)}\sqrt{\frac{n_b}{n_a}}\sqrt{\pi(2l+1)} (\mp p_x +i p_y) \\
&\int_{0}^{\alpha_\text{NA}} \,d\alpha \sin\alpha \sqrt{\cos\alpha} e^{- f^2 \sin^2 \alpha/\omega^2} e^{-ikz_0\cos\alpha} \left(\pi_l^1(\alpha)+\tau_l^1(\alpha)\right),
\end{split}
\end{equation}
while for $G^{TE}$ we obtain
\begin{equation}
\begin{split}
&G_{l,\pm 1}^{TE}=-\frac{i^l (k f e^{ikf})}{l(l+1)}\sqrt{\frac{n_b}{n_a}}\sqrt{\pi(2p+1)} (p_y \pm i p_x)\\
&\int_{0}^{\alpha_\text{NA}} \,d\alpha \sin\alpha \sqrt{\cos\alpha} e^{- f^2 \sin^2 \alpha/\omega^2} e^{-ikz_0\cos\alpha} \left(\pi_l^1(\alpha)+\tau_l^1(\alpha)\right),
\end{split}
\end{equation}

that can be summed up as follows:

\begin{equation}
G_{l,\pm 1}^{TM}=(\mp p_x +i p_y) G_l, \qquad G_{l,\pm 1}^{TE}=-(p_y \pm i p_x) G_l ,
\end{equation}

where

\begin{equation}
\begin{split}
&G_l=\frac{i^l (k f e^{ikf})}{l(l+1)}\sqrt{\frac{n_b}{n_a}}\sqrt{\pi(2l+1)} \\
&\int_{0}^{\alpha_\text{NA}} \,d\alpha \sin\alpha \sqrt{\cos\alpha} e^{- f^2 \sin^2 \alpha/\omega^2} e^{-ikz_0\cos\alpha} \left(\pi_l^1(\alpha)+\tau_l^1(\alpha)\right).
\end{split}
\end{equation}

\section{Optical Forces}

The optical force is derived from the conservation law of electromagnetic wave momentum. For a trapped spherical scatterer, the total electromagnetic force may be determined by applying Maxwell's stress tensor (MST) in the space surrounding the scatterer. A step-by-step application of the MST for an arbitrary incident wave and a dielectric sphere is presented. With the EM field described in terms of spherical harmonics, the observable effects of the force components that an EM wave exerts on the microsphere is given by the time average of the Maxwell stress tensor ($T_{ij}$), which for steady--state optical conditions is given in the traditional Minkowski form, defined below as in Ref.\cite{Jackson1999},

\begin{equation}\label{eq:MST}
\begin{split}
F_i&=\oint T_{ij} n_j \,dA\\
&=\frac{1}{2}\Re\oint \left[\epsilon E_i E_j^* + \mu H_i H_j^*-\frac{1}{2}\left(\epsilon \mathbf{E}\cdot\mathbf{E^*}+\mu \mathbf{H}\cdot\mathbf{H^*}\right)\delta_{ij}\right]n_j \,dA,
\end{split}
\end{equation}

where the EM fields are the ones outside the scatterer (incoming and scattered fields). We integrate over a spherical surface surrounding the scatterer, which can be chosen arbitrarily; hence, we choose one whose radius $kr\rightarrow\infty$. Where $n_j=\mathbf{\hat{r}}$, and because $dA=r^2 d\Omega$, only certain terms of the radial function expansion is needed. At infinity, the electromagnetic fields only contain transverse components, and therefore the first two terms of the right-hand side of Eq.(\ref{eq:MST}) do not contribute, resulting in

\begin{equation}\label{eq:StressTensor}
\mathbf{F}=-\frac{r^2}{4}\Re\int \left(\epsilon \mathbf{E}\cdot\mathbf{E^*}+\mu \mathbf{H}\cdot\mathbf{H^*}\right) \mathbf{\hat{r}} \,d\Omega.
\end{equation}

The EM fields involve spherical vector terms such as
\begin{equation}
\nabla\times(z_l(kr)\mathbf{X}_{lm}(\mathbf{\hat{r}}))=k  z_l'(kr)\left(\hat{r}\times\mathbf{X}_{lm}(\mathbf{\hat{r}})\right)+z_l(kr)\nabla\times\mathbf{X}_{lm}(\mathbf{\hat{r}}).
\end{equation}
Which for $kr\rightarrow\infty$ would result in
\begin{equation}
\nabla\times(z_l(kr)\mathbf{X}_{lm}(\mathbf{\hat{r}}))=k  z_l'(kr)\left(\hat{r}\times\mathbf{X}_{lm}(\mathbf{\hat{r}})\right).
\end{equation}

The asymptotic expansion of the spherical radial functions and their derivatives in this limit are
\begin{align}
j_l(kr)&=(-i)^{l+1}\frac{e^{ikr}}{2kr}+i^{l+1}\frac{e^{-ikr}}{2kr},\\
j_l'(kr)&=(-i)^{l}\frac{e^{ikr}}{2kr}+i^{l}\frac{e^{-ikr}}{2kr},\\
h_l^{(1)}(kr)&=(-i)^{l+1}\frac{e^{ikr}}{kr},\\
h_l'^{(1)}(kr)&=(-i)^{l}\frac{e^{ikr}}{kr}.
\end{align}

The total EM field outside the sphere and the sum of the incident and scattered fields are known in terms of the spherical vector harmonic. Inserting these fields into Eq.(\ref{eq:StressTensor}), and taking into account that the incident field contribution is removed as it is independent of the sphere position, the contribution of both fields to the force amounts to zero, i.e., all terms that do not contain $a_l$ and $b_l$.
The vectorial spherical harmonic products can be re-written as follows,

\begin{equation}
\begin{split}
&\left(\hat{r}\times\mathbf{X}_{lm}\right) \cdot \left(\hat{r}\times\mathbf{X}_{pq}^*\right) = \hat{r} \cdot \left[\mathbf{X}_{pq}^* \times \left( \hat{r}\times\mathbf{X}_{lm} \right) \right] =\\
&=\hat{r} \cdot \left[ \hat{r} \mathbf{X}_{lm} \cdot \mathbf{X}_{pq}^* - \mathbf{X}_{lm} \left( \hat{r} \cdot \mathbf{X}_{pq}^* \right)\right]= \mathbf{X}_{lm} \cdot \mathbf{X}_{pq}^*,
\end{split}
\end{equation}
and
\begin{equation}
\mathbf{X}_{pq}^* \cdot \left( \hat{r} \times \mathbf{X}_{lm}\right) = - \mathbf{X}_{pq}^* \cdot \left(\mathbf{X}_{lm}\times\hat{r}\right)=- \mathbf{X}_{lm}\cdot \left(\hat{r}\times\mathbf{X}_{pq}^*\right).
\end{equation}

These simplify the electric and magnetic terms to two components each. In this way, we can sum the contributions from the electric and magnetic terms, resulting in a force

\begin{equation}\label{eq:Force}
\begin{split}
&\mathbf{F}=\frac{\epsilon | E_0 |^2}{4 k^2} \Re \sum_{l=1}^\infty\sum_{m=-l}^{l} \sum_{p=1}^\infty\sum_{q=-p}^{p} \,  i^{p-l}\\
&\left\{ \left[ G_{lm}^{TM}G_{pq}^{TM*}(a_l+a_p^*-2a_la_p^*)+G_{lm}^{TE}G_{pq}^{TE*}(b_l+b_p^*-2b_lb_p^*)\right] \int \mathbf{X}_{lm} \cdot \mathbf{X}_{pq}^* \,\hat{r} \,d\Omega \right.\\
&\left. + \left[ G_{lm}^{TM}G_{pq}^{TE*}(a_l+b_p^*-2a_lb_p^*)-G_{lm}^{TE}G_{pq}^{TM*}(b_l+a_p^*-2b_la_p^*)\right] \int \mathbf{X}_{lm} \cdot \left(\hat{r}\times\mathbf{X}_{pq}^*\right)\,\hat{r}  \,d\Omega \right\}.
\end{split}
\end{equation}

The force is generally split into components, such that the explicit integrals over the solid angle are given in terms of

\begin{equation}
\begin{split}
&\int \sin\theta \mathbf{X}_{lm} \cdot \mathbf{X}_{pq}^* e^{i\phi} \,d\Omega \\
&= \frac{\delta_{q,m+1} \delta_{l,p+1}}{p+1}\sqrt{\frac{p(p+2)(p-m+1)(p-m)}{(2p+3)(2p+1)}} -\frac{\delta_{q,m+1} \delta_{l+1,p}}{l+1}\sqrt{\frac{l(l+2)(l+m+2)(l+m+1)}{(2l+3)(2l+1)}},
\end{split}
\end{equation}

\begin{equation}
\int \sin\theta \mathbf{X}_{lm} \cdot \left(\hat{r}\times\mathbf{X}_{pq}^*\right) e^{i\phi} \,d\Omega = \delta_{q,m+1} \delta_{l,p} \frac{i \sqrt{(l-m)(l+m+1)}}{l(l+1)},
\end{equation}

\begin{equation}
\begin{split}
&\int \cos\theta \mathbf{X}_{lm} \cdot \mathbf{X}_{pq}^* \,d\Omega \\
&= \frac{\delta_{q,m} \delta_{l,p+1}}{p+1} \sqrt{\frac{p(p+2)(p+m+1)(p-m+1)}{(2p+3)(2p+1)}}+\frac{\delta_{q,m} \delta_{l+1,p}}{l+1}\sqrt{\frac{l(l+2)(l+m+1)(l-m+1)}{(2l+3)(2l+1)}},
\end{split}
\end{equation}

\begin{equation}
\int \cos\theta \mathbf{X}_{lm} \cdot \left(\hat{r}\times\mathbf{X}_{pq}^*\right) \, d\Omega = \delta_{q,m} \delta_{l,p} \, \frac{im}{l(l+1)}.
\end{equation}

After substituting these integrals in Eq.(\ref{eq:Force}), we obtain after some simplification the transverse and longitudinal force components as follows:

\begin{equation}
\begin{split}
\begin{bmatrix}
           F_{x} \\
           F_{y}
\end{bmatrix}&=\frac{-\epsilon | E_0 |^2}{(2k)^2}
\begin{bmatrix}
           \Re \\
           \Im
\end{bmatrix}
\sum_{p=1}^\infty\sum_{q=-p}^{p} \frac{i}{p+1} \left\{ \sqrt{\frac{p(p+2)(p+q+2)(p+q+1)}{(2p+3)(2p+1)}} \right.\\
&\times \left[ A_p G_{p+1,-(q+1)}^{TM}G_{p,-q}^{TM*}+A_p^{*} G_{pq}^{TM}G_{p+1,q+1}^{TM*}+B_p G_{p+1,-(q+1)}^{TE}G_{p,-q}^{TE*}+B_p^{*} G_{pq}^{TE}G_{p+1,q+1}^{TE*} \right]\\
&\left.-\frac{\sqrt{(p-q)(p+q+1)}}{p} \left[ C_p G_{pq}^{TM}G_{p,q+1}^{TE*}-C_p^{*} G_{pq}^{TE}G_{p,q+1}^{TM*}\right]\right\},
\end{split}
\end{equation}

\begin{equation}
\begin{split}
F_{z}&= -\frac{\epsilon | E_0 |^2}{2 k^2} \Re \sum_{p=1}^\infty \sum_{q=-p}^{p} \frac{i}{p+1} \left\{ \sqrt{\frac{p(p+2)(p+q+1)(p-q+1)}{(2p+1)(2p+3)}} \right.\\
&\left.\times \left[A_p G_{p+1,q}^{TM} G_{pq}^{TM*} + B_p G_{p+1,q}^{TE}G_{pq}^{TE*}\right] -\frac{q}{p} C_p G_{pq}^{TM}G_{pq}^{TE*}\right\}.
\end{split}
\end{equation}

where the combination of Mie coefficients has been defined as the following

\begin{equation}
\begin{split}
A_{p}&=a_{p+1} + a_p^{*} - 2 a_{p+1} a_p^{*},\\
B_{p}&=b_{p+1} + b_p^{*} - 2 b_{p+1} b_p^{*},\\
C_{p}&=a_{p} + b_p^{*} - 2 a_{p} b_p^{*}.
\end{split}
\end{equation}

The force calculations, outlined step-by-step above, can easily be generalized for optical torques using the same stress tensor and guidelines presented in this tutorial.
This methodology was used to calculate the force in an optical force spectroscopy experiment \cite{TeseNeves2006}. In Ref.~\cite{Neves2006OL}, the figures show a comparison between the expansion method and the real beam. This model also explained the results from a double tweezers setup to perform an ultra--sensitive force measurement of the full optical force profile as a function of radial position and wavelength for both transverse electric (TE) and transverse magnetic (TM), modes excitation \cite{Neves2006OE}. Aberrations can also be considered, where the model successfully explained the ripple structure in the optical forces profile due to spherical aberrations due to refractive index mismatching \cite{Neves2007}.

\section{Final Remarks}

The infinite summation for the optical force may be of concern. Computationally, we truncate it to a particular value ($N_{max}$), because all other terms have a negligible contribution. This maximum summation term scales with the scatterer size (size parameter, $x$), just as in the determination of the scattering cross section, which uses the Wiscombe criterion \cite{Wiscombe1980}. By using a different floating-point variable for the numerical computation, a suitable truncation number can be chosen \cite{Neves2012}, by taking the closest integer of

\begin{equation}
N_{max}=0.76 \epsilon^{2/3} x^{1/3} - 4.1,
\end{equation}

where $10^{-\epsilon}$ is the relative truncation error, and $x$ is the size parameter defined in Eq.(\ref{eq:SizePar}). This is correct only for the total cross section, as has been emphasized recently \cite{Allardice2014}, or for EM fields within a region of space delimited by $k_2 r < N_{max}$. However, for specific fields instead of cross sections, a similar rule is applicable in which the maximum number of terms should be greater than the largest $k_2 r$ in the plot region.

For most applications, when the bead is near its equilibrium position in the optical trap, the trap stiffness ($\mathbf{\kappa}$) may suffice, which can be directly calculated from the optical force from Hooke's law,

\begin{equation}
\kappa_q = \left( \frac{\partial F_q}{\partial q}\right)_{r=r_{eq}},
\end{equation}

where $q=x,y,\text{ or }z$, and is evaluated at the equilibrium position ($r_{eq}$).
Many refinements have been made to render the approach described here more complete. One can take into account the spherical aberrations produced by the refractive index mismatch at the coverslip interface \cite{Torok1995}, whose consequences on optical forces are known \cite{Rohrbach2002,Fallman2003,Vermeulen2006}. The appropriate aberrated beam coefficients can be determined in a similar way \cite{Neves2007}, making the optical force more realistic when trapping close to an interface. 

One of the main efficiencies in determining optical forces with this procedure, are the analytically determined BSCs at an arbitrary position with respect to the microsphere, without the need of introducing computational error from translation operations. The present method allows to obtain the resultant force on a spherical dielectric by displacing the particle with respect to beam focus, and hence a full force profile can be obtained.

The precise determination of the BSC for a highly focused beam has been validated experimentally, making it a robust strategy for BSC determination. One of these validations was the measurement of a complete force profile for microspheres of different sizes, and the effects of coupling the beam polarization to different TM and TE modes separately \cite{Neves2006OE}. Recently, this force profile or force profile mapping measurement and comparison have been performed with an improved setup \cite{Jahnel2011,Ignacio2012,Ignacio2013}. The analytical BSCs was also validated independently in a single particle photo-thermal experiment \cite{Selmke2012}.

Another method to analytically determine the BSCs was derived by Moreira \textit{et. al} using the Fourier transform of the angular momentum operator \cite{Moreira2010,Moreira2016}. In this work, the BSCs are for the first time shown to be independent of the radial coordinate if the beam model exactly satisfies Maxwell's wave equations. Several, analytical BSC for different beam modes, such as Bessel beams, circular, and rectangular wave-guided beams were likewise derived. This new method also applies to plane waves and highly focused beams.

As a final contribution of this study, the computer codes generated as a result of all of the work presented in this paper are released and are freely available, also allowing for contributions from the community\cite{GitHub}.

\section{Summary}

We presented an overview of the approach to model optical forces occurring in optical tweezers. The optical forces are determined by the Maxwell stress tensor approach and expressed in terms of the beam shape coefficients from the Generalized Lorenz-Mie Theory. A robust beam model is employed for the tightly focused Gaussian beam with a high NA objective lens in terms of the diffraction integrals presented by Richards and Wolf and expressed by the angular spectrum representation. From this beam, the beam shape coefficients are determined by analytically integrating under the solid angle.

This work allows for accurate and precise computational modeling to quantitatively compare the theoretical results with experimental outcomes. Using the analytical expressions presented in this work requires low computational complexity. We hope that this work can be of help to help young researchers in the field and provide them with a more solid background in optical forces theory, stimulating further development for trapping applications.

\section*{Funding}
This work was partially funded by the São Paulo Research Foundation (FAPESP) (2017/13543-3).\\
\bibliography{sample}






\end{document}